\documentclass[manuscript,nonacm]{acmart}
\AtBeginDocument{%
  \providecommand\BibTeX{{%
    \normalfont B\kern-0.5em{\scshape i\kern-0.25em b}\kern-0.8em\TeX}}}

\setcopyright{acmcopyright}
\copyrightyear{2022}
\acmYear{2022}
\acmDOI{}

\acmConference{2022}
\acmBooktitle{}
\acmPrice{}
\acmISBN{}

\usepackage{multirow, makecell}
\begin{document}


\title{Identifying Ethical Issues in AI Partners in Human-AI Co-Creation}
\author{Jeba Rezwana}
\email{jrezwana@uncc.edu}
\author{Mary Lou Maher}
\email{m.maher@uncc.edu}
\affiliation{%
  \institution{University of North Carolina at Charlotte}
  \country{USA}
  \postcode{}
}

\renewcommand{\shortauthors}{Rezwana and Maher}


\begin{abstract}
Human-AI co-creativity involves humans and AI collaborating on a shared creative product as partners. In many existing co-creative systems, users communicate with the AI using buttons or sliders. However, typically, the AI in co-creative systems cannot communicate back to humans, limiting their potential to be perceived as partners. This paper starts with an overview of a comparative study with 38 participants to explore the impact of AI-to-human communication on user perception and engagement in co-creative systems and the results show improved collaborative experience and user engagement with the system incorporating AI-to-human communication. The results also demonstrate that users perceive co-creative AI as more reliable, personal and intelligent when it can communicate with the users. The results indicate a need to identify potential ethical issues from an engaging communicating co-creative AI. Later in the paper, we present some potential ethical issues in human-AI co-creation and propose to use participatory design fiction as the research methodology to investigate the ethical issues associated with a co-creative AI that communicates with users.
\end{abstract}

\ccsdesc[500]{Human-centered computing~User studies}

\keywords{Co-creativity, Interaction design, Ethical AI, Design Fiction, Human-AI Co-Creation, Ethical Issues}


\maketitle

\vspace{-0.15cm}
\section{Introduction}
Human-AI co-creativity, a subfield of computational creativity, involves both humans and AI collaborating on a shared creative product\cite{davis2013human}. Human-AI co-creativity is an important area of study given that AI is being used increasingly in collaborative spaces, including AI in collaborative music \cite{hoffman2010shimon}, collaborative design \cite{karimi2018evaluating} or even in hospitals as a virtual nurse \cite{crowder2020human}. This emerging research field has a number of open questions, including how to design co-creative systems \cite{kantosalo2014isolation, davis2016empirically}. The technological ability of AI alone does not ensure a positive collaborative experience of users with the AI \cite{louie2020novice}. Current human-centered AI (HAI) research emphasizes that the next frontier of AI is not just technological but also humanistic and ethical: AI is to enhance humans rather than replace them \cite{xu2019toward}. Only humans can communicate to AI in many existing co-creative systems, usually with buttons, sliders, or other UI components. However, in many existing co-creative systems, AI can not communicate back to users \cite{10.1145/3519026} which is essential for a co-creative AI to be considered as a partner. In a human collaboration, collaborators communicate to provide feedback and convey information \cite{gutwin1996workspace}. We investigated the impact of incorporating AI-to-human communication via speech, text and visuals (a virtual AI character/avatar) along with human-to-AI communication on \emph{collaborative experience, user engagement} and \emph{user perception} of co-creative AI in human-AI co-creativity. 

We conducted a user study with 38 participants with two high-fidelity interactive prototypes of a co-creative system, with and without AI-to-human communication. The results demonstrate that two-way communication, including AI-to-human communication along with human-to-AI communication, improves the collaborative experience and user engagement as the co-creative AI is perceived as a collaborative partner. Users perceive the co-creative AI that communicates with users as more intelligent, personal, and reliable. Our results present ethical concerns and potential ethical issues in an engaging co-creative AI with human-like attributes. Ethical issues associated with autonomous AI have been widely explored, although ethical human-AI co-creation is rarely explored in the literature. We present some ethical questions and issues in the context of co-creative AI based on the user study results and the literature. We propose using design fiction as a research tool to identify ethical issues associated with co-creative AI that engages users with human-like attributes. 
\vspace{-0.1cm}

\section{User Study and Results}
We recruited 38 participants for the study, 19 males and 19 females, which involved user interaction with two prototypes of a co-creative system that contributes design inspirations during a design task. One prototype utilizes only human-to-AI interaction using buttons (baseline). Another prototype uses two-way communication between humans and AI, including AI-to-human communication via speech, text and an AI avatar. The AI ability was controlled for the study as both systems have the same ability in terms of what they can do. We used a within-subject design method as each participant tested both conditions and we counterbalanced the order of the conditions. The design tasks for the two conditions were- design a futuristic chair for gamers/ shopping cart specifically designed for the elderly. After each design task, the participants completed a survey to evaluate the system and reflect on the co-creation experience. Finally, the study ended with a follow-up semi-structured interview to collect qualitative data about the user perception of the AI, and overall experience with the AI. The study did not require participants to have drawing/sketching skills. 

We conducted a statistical analysis of the survey data for evaluating the co-creation experience. We used T-tests comparing the effect of each condition on our outcome variables and conducted a thematic analysis on our semi-structured interviews. Our results show that two-way communication, including AI-to-human communication, improves the collaborative experience. The T-test revealed a significant difference (p=0.002<0.05) between the collaboration scores of the two prototypes as participants scored their collaborative experience with the communicating AI significantly higher than the baseline AI. Additionally, results from the thematic analysis show that most of the participants (n=30) reported that the experience with the communicating AI felt more like collaborating with a partner as there was good communication, unlike the baseline AI. Most participants liked the aspect that the AI speaks to them and asks for user feedback on its contributions. Participants also liked the affective characteristics of the AI avatar, for example, visually being sad when users did not like its inspirations or being happy when they liked an inspiration. The results from thematic analysis demonstrated that participants engaged more with the communicating AI than with the baseline AI. Most participants enjoyed using the communicating AI more than the baseline AI as communicating AI was more interactive and kept them attentive. Participants reported a sense of awareness of another ‘being’ while collaborating with the communicating AI, which helped them be more attentive. However, survey results did not demonstrate any significant difference in engagement factor between the two prototypes.

The communicating AI was perceived as a more intelligent, personal and reliable AI. Many participants perceived that communicating AI helped and guided them more than baseline AI. P7 shared how the communicating AI was more helpful and reliable, “\emph{It was definitely, more helpful in allowing me and in guiding me to what I wanted to draw than the other one.}” Participants felt like the communicating AI understood their needs better than the baseline. “\emph{The first one} (communicating AI)\emph{ was more in sync with my thoughts and was more AI-ish}”, said P4. Participants also perceived the communicating AI as more personal as they felt a connection with it. Regarding this P12 said, “\emph{I think adding the simple feature, like speaking to you and listening to you, made it more personal.}” Some participants wanted to talk back to the AI as it seemed more fun, personal and reliable. P19 said, “\emph{ It would be cool if I could talk to the AI}”. Findings show that the perceived creativity of the final design differs between the prototypes. Most participants preferred the final design created with the communicating AI as more creative. These findings show how changing one aspect of interaction design changes user perception and engagement of a co-creative AI even though there is no change in the AI ability.

\vspace{-0.25cm}
\section{Potential Ethical Issues in Human-AI Co-Creativity}
Increased user engagement and perceived reliability accomplished only through interface design may appear to be a good outcome, but it also raises ethical concerns. People's perceptions about AI's trustworthiness and connection with AI have an impact on their decisions and actions. Because AI optimization can evolve quickly and unexpectedly, the challenge of value alignment arises to ensure that AI's goals and behaviors align with human values and goals \cite{wallach2008moral, russell2015ethics}. Ethically aligned design is a must for human-centered AI solutions that avoid discrimination, maintain fairness, and justice \cite{xu2019toward}. When AI is incorporated in social entities and interacting with us, questions of values and ethics become even more urgent \cite{liao2016can}. Because co-creative AI represent forms of autonomous technology that can affect human partners, it is essential to anticipate ethical issues and address them during all steps of design \cite{muller2017exploring}. In a human-AI co-creative system, AI not only categorizes, evaluates and interprets the data but also creates data itself. In a co-creative system, the role of AI changes from a lone decision-maker to a more complex one depending on the partnership. The complex interaction and partnership make it difficult to answer `who owns the product in a human-AI co-creation? The human or the AI'? The user study results raise questions like- do increased engagement and human-like attributes of AI in human-AI co-creation increase ethical concerns? The participants in the user study did not know that the AI ability was the same for both prototypes and perceived the communicating AI as more intelligent and reliable. So another question emerges - will being transparent to users about the AI ability produce the same effect? We need to investigate these questions in the context of co-creative AI. However, existing ethical guidelines for responsible AI technologies will help identify the initial measures for designing ethical co-creative AI as they represent a type of autonomous AI. 

Amershi et al. proposed 18 generally applicable design guidelines for human-AI interaction for AI technologies which include transparency and mitigating unfair social stereotypes and biases \cite{amershi2019guidelines}. Ruane et al. asserted that the ethical issues emerge from the traditional black-box model of AI and suggested transparency and explainability for the AI's motivation and behavior \cite{ruane2019conversational}. The communication and dialogue design between AI and humans impacts users' inclination to self-disclose unintentional data \cite{ruane2019conversational}. Self-disclosure may be encouraged to gather data to improve user experience \cite{saffarizadeh2017conversational}. However, users may not be aware of how much information they have shared or what personal data can be inferred from the interaction with the AI. Users may also be unaware of how the system operates on a technical level in terms of data processing and storage \cite{luger2016like}. For these reasons, the unique context of co-creative AI that can communicate to users concerning user privacy should be considered while devising the ethical guidelines for co-creative AI. It is also important to consider the impact of the co-creative AI persona on the human-AI partnership and determine if the AI persona is encouraging behavior that may be harmful. AI agent persona can also inadvertently reinforce harmful stereotypes \cite{ruane2019conversational}. AI agent persona may be more explicit as the level of embodiment increases. Besides, to make an AI free of biases, it is important to consider what kind of data an AI has access to train on to generate its content and make a decision, and what kind of information the AI should show to the users. 

\vspace{-0.15cm}
\section{Design Fictions as a Research Method to Investigate Ethical Issues in Human-AI Co-Creation}
To identify ethical issues in terms of user perspective, user accounts of technologies they envision and values that co-creative AI implicates need to be investigated. Understanding humans in a design area where they may not have lived but have had some experiences through popular culture is a major challenge \cite{muller2017exploring}. Such experiences unavoidably shape user needs and values when interacting with AI goods, but they are too vague for developing systems \cite{muller2017exploring}. Human-AI co-creativity research is still formative and might still be alien or abstract to ordinary people. In this context, we need methods that are more likely to tell us about what we don't know about the unknown future of co-creative systems. Muller and Liao proposed design fictions to restore the future users to a central position in anticipating, designing, and evaluating future intelligent systems \cite{muller2017exploring}. They proposed this method to reflect the interests and the values of the future users through working with them to design the ethics and values of intelligent systems. Design Fictions (DF) is a relatively new research and design practice term that Bruce Sterling coined in Shaping Things \cite{sterling2005shaping} where he describes DF as the “deliberate use of diegetic prototypes to suspend disbelief about change”. Design fiction depicts a future technology through the world of stories, and users express their own accounts of the technologies they envision, as well as the values that those future technologies implicate \cite{muller2017exploring}. DF have been used to reveal values associated with new technologies \cite{brown2016ikea, dourish2014resistance, tanenbaum2016limits} and to open a space for diverse speculations about future technologies \cite{blythe2014research}. We will use DF as a participatory method to describe an engaging futuristic co-creative AI using the world of fiction (narrative or videos) so that humans can react to it and modify it to express their values, concerns and expectations. We will identify a design fiction from the literature and adapt it to fit it in the context of futuristic co-creative AI. 

Fictions also raise concerns about autonomous AI and robots with human-like attributes and intelligence; for example, in the film Ex Machina, we witness a human-like robot obtain human conscience by observing and learning from a human trainer, then faking affection for him and eventually killing him. In the film Her, a guy develops feelings for his operating system, which has a persona and human-like attributes. In 2001: A Space Odyssey, a narrative science fiction classic by Arthur C. Clarke, a supercomputer with conversational AI, HAL, killed the spaceship crew so that it could achieve its goal, which was actually set by humans. Another similar example is Cat Pictures Please by Naomi Kritzer, a narrative short fiction in which an autonomous AI manipulates the users' lives in an unethical manner such that their lives are flawless on its own terms. All of these examples demonstrate ethical issues with independent AI and robots. However, we rarely witness fiction in the form of movies or narratives regarding ethical dilemmas emerging from a co-creative AI partner that directly collaborates with humans and generates new data. Writing and encapsulating a design fiction in the context of co-creativity from scratch is difficult and time-consuming. Therefore, we will adopt a design fiction from the literature and adapt it to fit the context of human-AI co-creation. We will utilize participatory DF with users to investigate ethical issues associated with a co-creative AI along with their expectations of a co-creative AI. In the literature, multiple methods have been offered to practice DF as a research methodology \cite{10.1145/2513506.2513531, grand2010design} and we will use these methods as guides to practice DF to investigate the ethical issues associated with human-AI co-creation.

\vspace{-0.25cm}
\section{Conclusions}
This paper presents an overview of results and insights from a user study investigating the impact of two-way communication between co-creative AI and humans, including AI-to-human communication on user engagement, collaborative experience, and user perception. The results indicated further investigation for ethical issues and user accounts on designing human-centered fair co-creative AI. We present some ethical questions and issues in the context of co-creative AI based on the literature and the results from the user study. We conclude the paper by stating the reasons for choosing design fiction as a methodology for investigating the ethical issues in human-AI co-creativity and our method to practice design fictions.

\bibliographystyle{ACM-Reference-Format}
\bibliography{sample-base}


\end{document}